\documentclass[conference]{IEEEtran}

\usepackage{cite}
\usepackage{graphicx}
\usepackage{amssymb,hhline,enumerate,dsfont}
\usepackage{amsmath}
\usepackage{array}
\usepackage{amsmath,url}
\usepackage{psfrag}
\usepackage[usenames,dvipsnames]{pstricks}
\usepackage{epsfig}
\usepackage{pst-grad} 
\usepackage{pst-plot} 

\newtheorem{thm}{Theorem}

\IEEEoverridecommandlockouts
\begin{document}
%
\title{\vspace{-0.08in}Overhead-Optimized Gamma Network Codes}
\author{\IEEEauthorblockN{Kaveh Mahdaviani}
\IEEEauthorblockA{ECE Department, University of Toronto, Canada\\
kaveh@comm.utoronto.ca}
\and
\IEEEauthorblockN{Raman Yazdani, Masoud Ardakani}
\IEEEauthorblockA{ECE Department, University of Alberta, Canada\\
\{yazdani,ardakani\}@ece.ualberta.ca}}


%

\maketitle

\begin{abstract}
We design a network coding scheme with minimum reception overhead and linear encoding/decoding complexity.
\end{abstract}


%
\section{Introduction} \label{sec:introduction}
In sparse random linear network coding (SRLNC), a block of $K$ source packets is first divided into $n$ generations each having $g$ packets. At the source, each output packet is independently generated as a linear combination of the packets inside a randomly selected generation with coefficients selected uniformly at random from $\mathrm{GF}(q)$. Random selection of generations for encoding is performed to avoid the large number of control messages required for mitigating the problem of rare blocks and block reconciliation \cite{Efficient_Methods}. The intermediate nodes recode their received packets in the same fashion by producing random linear combinations of them without mixing packets from different generations. The receiver, collects enough linear combinations of source packets by receiving $r$ encoded packets to recover the source packets.

While SRLNC has linear encoding/decoding complexity, due to its similarity to the coupon collector's problem, its reception overhead $(r-K)/K$ scales logarithmically in $K$ \cite{RandomAnnex}. To mitigate this effect, one idea is to impose some dependency among the packets in different generations to help recovery of rare generations at the receiver. To this end, overlapping generations has been introduced \cite{Silva_Overlapping}. The efficiency of overlapping generations and the best pattern for overlap, however, have not been fully investigated.


Gamma network codes \cite{Mahdaviani12} impose dependencies among source packets using a linear pre-code of rate $1-\delta$, and a linear outer code of rate $R$ (giving rise to a block of $N$ packets) before splitting them into distinct generations and performing SRLNC. The outer code is characterized by a polynomial $P(x)=\sum_{i=2}^{D}p_ix^i$ where $p_{i}$ is the probability that a randomly selected check equation of the outer code involves packets from $i$ generations. The decoder then iterates between outer code and  SRLNC decoding and accomplishes the decoding through pre-code decoding. The encoding/decoding complexity of these codes is linear in terms of $K$. 

One advantage of Gamma network codes is that their performance is analytically tractable. In this work, we present an asymptotic analysis and use it to design overhead-optimized Gamma network codes.


\section{Asymptotic Analysis and Numerical Results} \label{sec:analysis}
We study the average performance of Gamma network codes under an asymptotic length assumption and large $q$.

\begin{thm}\label{convergence}
For a Gamma network code with a linear random outer code of rate $R$ and check degree distribution $P(x)$, if for some $0<x_0<1$ and $0<\delta<1$
\begin{equation*}\label{eq:convergence_condition}
x<1-\frac{\Gamma_g(r_{0}+g(1-R)P'(x))}{(g-1)!},~~\forall x\in (x_{0},1-\delta),
\end{equation*}
where $\Gamma_{\alpha}(x)=(\alpha-1)!e^{-x}\sum_{i=0}^{\alpha}{x^i/i!}$ and $r_0=\Gamma_g^{-1}\left((g-1)!(1-x_0)\right)$, then the Gamma network code can asymptotically recover all of the source packets using a linear capacity-achieving pre-code of rate $1-\delta$ with an overall average reception overhead of $\epsilon = r_0/(g(1-\delta)R)-1$.\hfill$\square$
\end{thm}

Using the above results and numerical optimization, we find a combination of the outer code and pre-code parameters ($R$, $P(x)$, $\delta$) giving the minimum reception overhead. For example, for $g=25$ and $D=15$, Gamma network codes is found with asymptotic average reception overhead as low as $\epsilon=2.75\%$. This further decreases to $\epsilon=1.92\%$ if $g=75$.

The above figure compares the performance of finite-length Gamma network codes constructed based on such optimized parameters with other existing SRLNC schemes \cite{Silva_Overlapping,Efficient_Methods,RandomAnnex}, assuming $q=2$ and $g=25$. Our pre-code is a right-regular LDPC code. It is evident that optimized Gamma network codes outperform the other schemes. Average reception overhead as low as $8.75\%$ is achievable when $N=16750$.


\begin{figure}
\centering
\includegraphics[width=0.92\columnwidth]{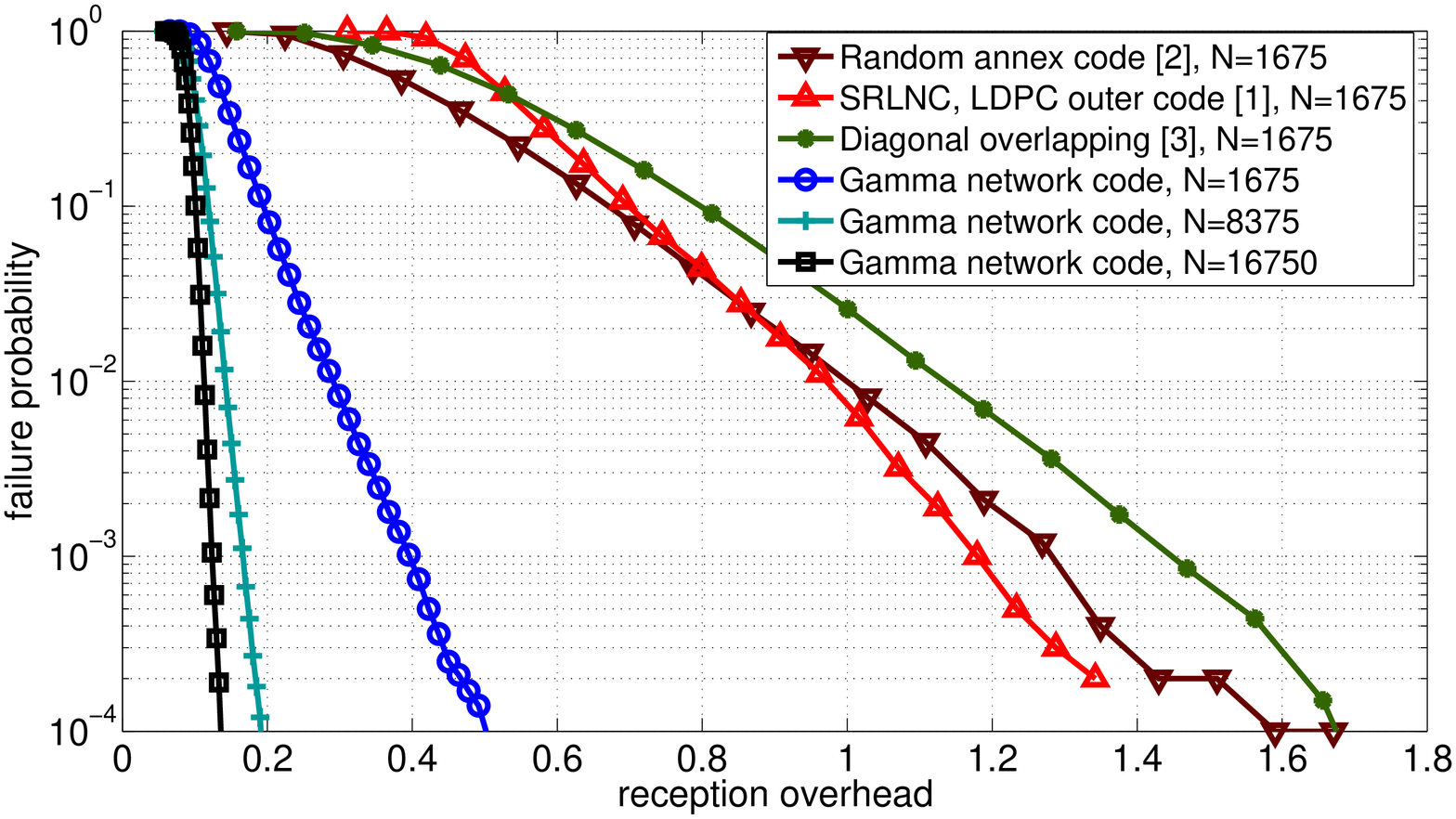}
\label{fig:overhead1}
\end{figure}

\bibliographystyle{IEEEtran}
\bibliography{IEEEabrv,Gamma_bib}

\end{document}